\documentstyle[psfig]{l-aa}

\begin{document}

\thesaurus{1(11.06.2; 11.16.1; 11.19.2; 11.19.6; 11.19.7)}

\title{{\it Letter to the Editor} \\
The shape of galaxy disks: how the scale height increases with
galactocentric distance\thanks{Based on observations obtained at the
European Southern Observatory, La Silla, Chile}}

\author{R.~de Grijs \and R.F.~Peletier}

\institute{Kapteyn Astronomical Institute, University of Groningen,
P.O. Box 800, 9700 AV Groningen, The Netherlands}

\offprints{R. de Grijs}

\date{Received: November 5, 1996; accepted: February 21, 1997}

\maketitle

\markboth{R. de Grijs and R.F. Peletier: The shape of galaxy disks}{}

\begin{abstract} 
We present the results of a detailed study of vertical surface
brightness profiles of edge-on disk galaxies.  Although the exponential
disk scale height is constant to first order approximation, we show that
for the large majority of galaxies in our sample, the scale height
increases with distance along the major axis.  The effect is strongest
for early-type galaxies, where the increase of the scale height can be
as much as a factor of 1.5 per scalelength, but is almost 0 for the
latest-type galaxies.  The effect can be understood if early-type disk
galaxies have thick disks with both scale lengths and scale heights
larger than those of the dominant disk component.  Its origin appears to
be linked to the processes that have formed the thick disk. 
\end{abstract}

\section{Disk vertical scale parameters}

The behaviour of the thickness of galactic disks is of major importance
for understanding the evolution of disk galaxies. Since the exponential
vertical scale height can be related directly to the vertical velocity
dispersion in (isothermal) disks (e.g., Bahcall \& Casertano, 1984), we
may be able to constrain models that describe the dynamical heating
mechanisms of galactic disks by studying this parameter. 

Van der Kruit \& Searle (1981a,b; 1982) found, for their sample of
edge-on spirals, that the vertical scale parameter, $z_0$ = 2 $h_z$, is
in good approximation independent of position along the major axis. 
Later studies of NGC 891 (Kylafis \& Bahcall, 1987) and NGC 5907
(Barnaby \& Thronson, Jr., 1992) have confirmed this result. From
two-dimensional modeling of a sample of 10 edge-on spiral and
lenticular galaxies, Shaw \& Gilmore (1990) found that the radial
variation of scale heights is typically within $\pm$ 3\% of the derived
mean for the main disk component, with no obvious dependence on colour
or model type adopted. 

\subsection{Why should the scale height be constant?}

Van der Kruit \& Searle (1982), following a suggestion by Fall, pointed
out that the scale height would be constant during the secular evolution
of a galactic disk if:
\begin{itemize} 
\item the disk is continuously heated by, e.g., the random acceleration
of the disk stars by giant molecular clouds or spiral structure (Spitzer
\& Schwarzschild, 1951). 
\item at all times the star formation rate is proportional to the
surface number density of the giant molecular clouds.
\end{itemize}

In this case the radial distribution of both the vertical velocity
dispersion and the surface mass density of the luminous matter would be
determined by the radial distribution of the giant molecular clouds. 
Since the vertical velocity dispersion and surface mass density possess
similar radial distributions, their ratio has no radial dependence. The
scale height also will not exhibit a dependence on radial distance,
since it is proportional to this ratio (van der Kruit \& Searle, 1981a;
van der Kruit, 1988). 

However, at present there is no satisfactory explanation as to how disk
heating, the rate of which must vary greatly with radius from the
observed distribution of molecular clouds in our own and other galaxies,
can naturally lead to this result. 

Alternatively, the so-called ``self regulation'' mechanism for galaxy
disks was used to explain the constancy of the vertical scale parameter
(see Bertin \& Lin, 1987). This mechanism, which describes the disk
evolution, leads to a more or less constant value for Toomre's (1964)
$Q$ parameter as a function of radius. Combined with an exponentially
decreasing velocity dispersion, as found in exponential disks, this
leads to a constant scale height (Bottema, 1993). 

However, on closer inspection the constancy of the scale height seems to
lose strength in the (radially) outer parts in a number of galaxies
studied. For our Galaxy, Kent et al. (1991) find indications that the
scale height increases linearly with radius from a minimum radius
outwards. The improvement over models with a fixed scale height is
significant, although not dramatic. 

\subsection{A statistically complete sample}

In order to be able to draw conclusions based on statistical
considerations rather than on individual cases we need a large,
homogeneous data set, selected in such a way that unwanted selection
biases are avoided. 

A diameter-limited sample, taken from the Surface Photometry Catalogue
of the ESO-Uppsala Galaxies (ESO-LV, Lauberts \& Valentijn 1989) was
used. 

We selected all non-interacting galaxies for which either the blue
angular diameter (at a surface brightness level of $\mu_B = 25$ mag
arcsec$^{-1}$), $D_{25}^B$, as measured by an automatic ellipse fitting
routine, or the original visual diameter, $D_{vis}$, is greater than
$2.'2$. From $V/V_{max}$ completeness tests (e.g., Davies, 1990) it
follows that the ESO-LV is statistically complete down to blue angular
diameters, $D_{25}^B$ and $D_{vis}$ combined, of $1.'0$. 

To the diameter-limited sample obtained we applied an inclination
selection criterion. As van der Kruit \& Searle (1981a) showed, for
systems having inclinations $i \ge 86^\circ$ the slopes of both the
radial and the vertical structures are independent of inclination. 
Therefore we defined our lower inclination limit to be 87$^\circ$. 
Inclinations were determined following de Grijs \& van der Kruit
(1996). 

For the observed sample the mean $V/V_{max}$ value results in $0.473 \pm
0.041$, so that we conclude that it is at least statistically complete. 

From the total sample of 93 edge-on galaxies we randomly selected 48,
which were observed in optical passbands ($B$, $V$, $R$, and $I$) using
both the Danish 1.54m and the Dutch 0.92m telescopes of the European
Southern Observatory at La Silla, Chile, from 1993 to 1996. Both
telescopes were used in direct imaging mode, at prime focus. 

\section{Results}

We extracted vertical luminosity profiles at a number of positions along
the major axes of the sample galaxies. To do this, the galaxies were binned
semi-logarithmically both radially and vertically to retain an
approximately constant overall S/N ratio in the resulting vertical
profiles and to be able to follow the profiles further out.

In the analysis of the individual profiles, we distinguished between the
different sides of the galaxy planes, to avoid possible dust
contamination in the case of not perfectly edge-on orientations. 

Both van der Kruit (1988) and de Grijs \& van der Kruit (1996), have
shown that the vertical profiles are more peaked than expected from an
isothermal distribution, but less peaked than exponential, probably
close to a sech({\it z}) distribution. The analysis shows that we can
approximate the profile's shape with an exponential distribution at {\it
z}-heights greater than 1.5 scale heights. We fit the vertical profiles
out to 4 scale heights, thereby taking into account the possible
presence of underlying thick disk components. Comparison for 24
galaxies with our unpublished $K'$-band images showed that in this
vertical region the contamination of the stellar light by dust
extinction can also be considered to be negligible. 

\begin{figure}
\vspace{-1cm}
\psfig{figure=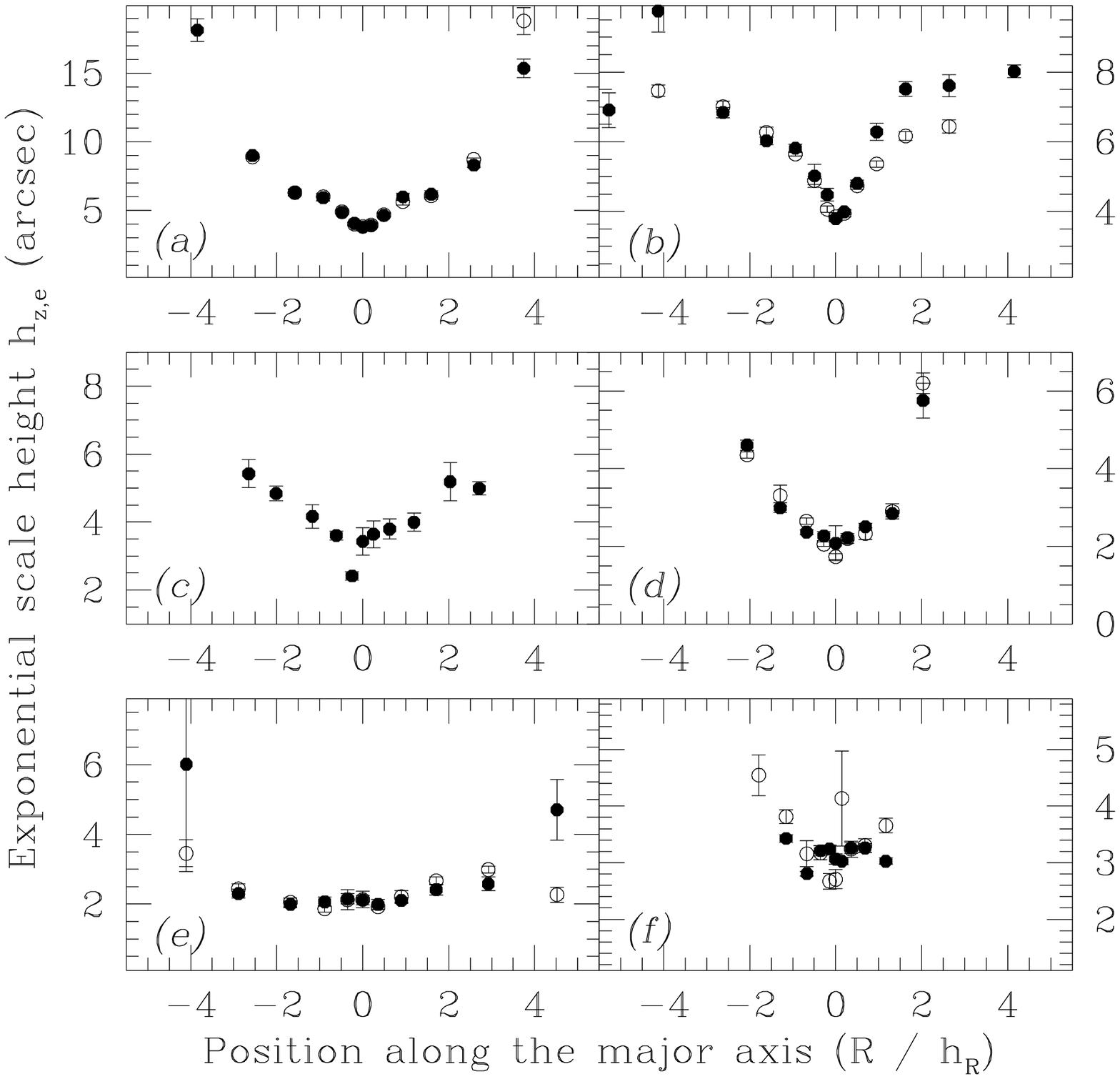,width=9cm}
\caption{}{\label{voorbeeld.fig}Examples of the {\it I}-band scale
height behaviour as a function of galactocentric distance for {\it (a)}
ESO 358G-29 (T = $-1.6$), {\it (b)} ESO 311G-12 (T = 0.0), {\it (c)} ESO
315G-20 (T = 1.0), {\it (d)} ESO 322G-87 (T = 3.4), {\it (e)} ESO
435G-50 (T = 5.0), and {\it (f)} ESO 505G-03 (T = 7.7). Open and closed
symbols represent data taken on both sides of the galaxy planes.}
\end{figure}

We note that the scale height seems to increase with radius along the
major axis, depending on galaxy type. As an example, in Fig. 
\ref{voorbeeld.fig} we show representative scale height distributions as
a function of position along the galaxies' major axes for galaxies
representing various types. 

\begin{figure}
\vspace{-4.2cm}
\psfig{figure=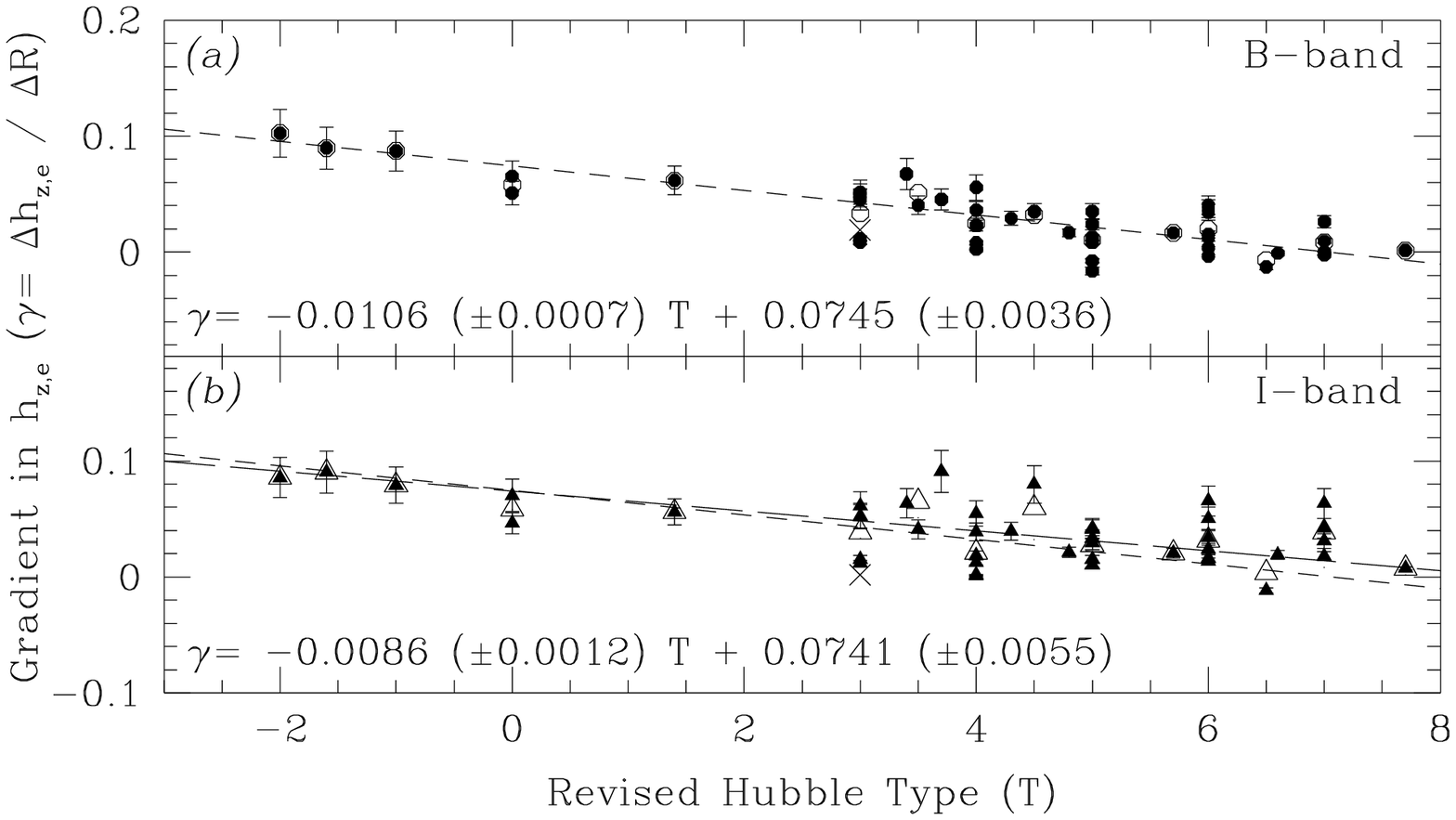,width=9cm}
\caption{}{\label{slopes.fig}Disk scale height gradients as a function
of revised Hubble type. The closed symbols represent the data; the open
symbols are type-averaged data points used to determine the relationship
described in the text. The results obtained for NGC 891 are shown by
crosses ($\times$). For comparison, the best fit obtained for the {\it
B}-band data is also shown in the {\it I}-band figure}
\end{figure}

\begin{figure}
\vspace{-0.5cm}
\psfig{figure=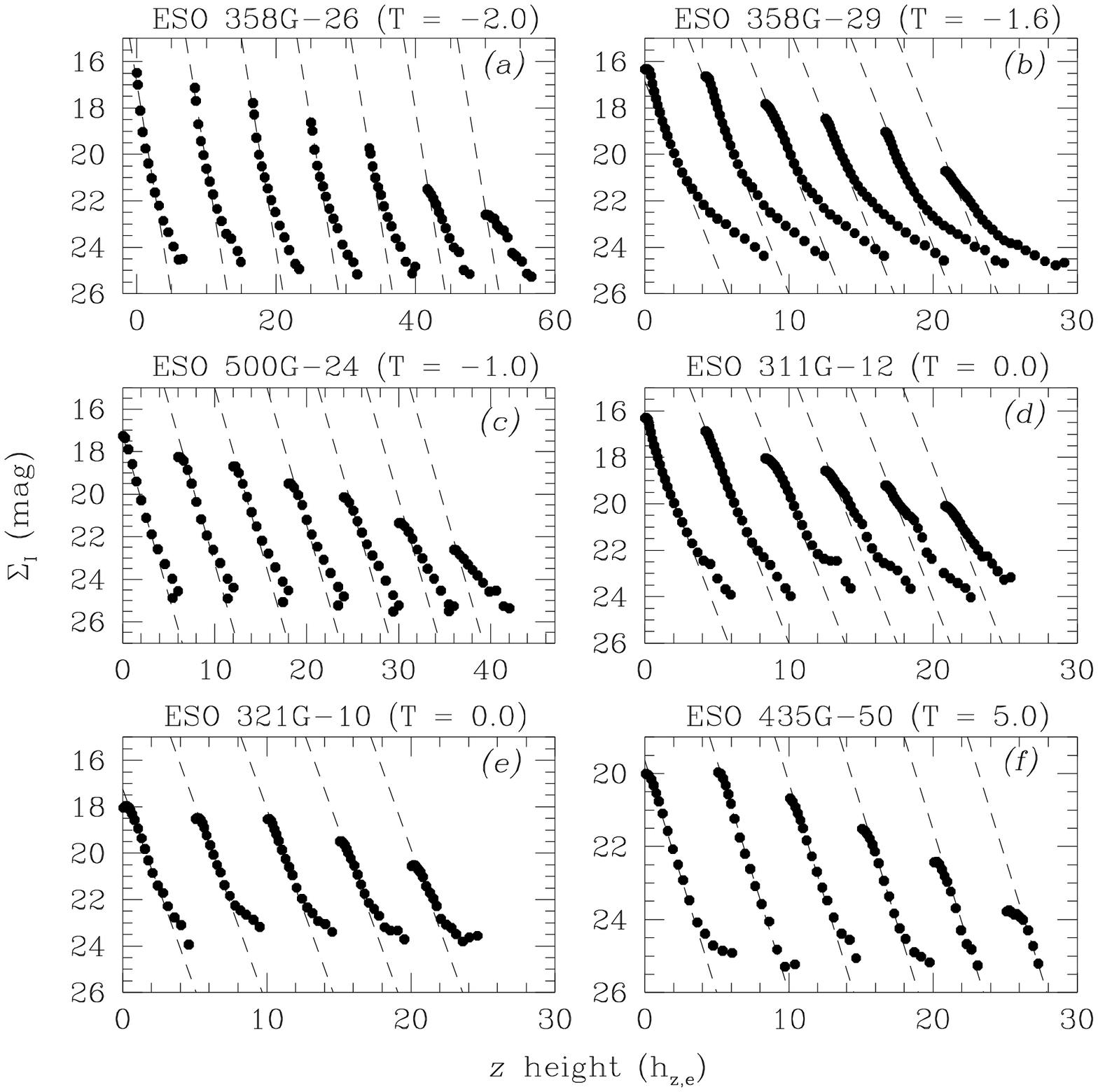,width=9cm}
\caption{}{\label{consthz.fig}{\it (a)} -- {\it (f)} Examples of
increasing scale height with galactocentric distance, compared to fixed
scale height models. The fitting of the slopes was done between 1.5 and
4 scale heights. From left to right the profiles that are shown were
extracted at logarithmically increasing galactocentric distances}
\end{figure}

We fit the scale height distribution as a function of position along the
major axis for the total range of data points, between 1.5 and 3.5, and
between 2.0 and 4.0 {\it I}-band scale lengths. The results obtained
for the fitting range between 2.0 and 4.0 scale lengths do not differ
substantially from the fitting range between 1.5 and 3.5 scale lengths,
whereas the scatter increases if we take all data points, including
those dominated by the bulge light contribution. 

Fig. \ref{slopes.fig} shows the results for both the {\it B} and the
{\it I}-band data, for the fitting range between 1.5 and 3.5 scale
lengths. We see that they do not differ significantly. This means that
the effect is intrinsic to the old disk, and not affected by, e.g., dust
or a young stellar population. Especially for the earlier-type galaxies
of our sample the observed relationship might be due to a non-negligible
bulge contribution. To discard this possibility, we varied our radial
fitting range. We estimated the bulge influence by applying a
two-dimensional bulge-disk decomposition to the data. For this sample,
the bulge contribution can be assumed to be negligible (i.e. less than
5\% of the total light) outside 1.8 $\pm$ 0.3 {\it I}-band scale
lengths, depending on galaxy type. 

To test our method, we applied the same procedure to the data of van der
Kruit \& Searle (1981b) for the edge-on galaxy NGC 891, and found that
the slopes obtained for NGC 891 fall in the same range as our more
recent data. NGC 891 is also included in Fig. \ref{slopes.fig}. 

To understand better the nature of this effect we show for a few
representative galaxies vertical profiles at various radial distances
from the center in Fig. \ref{consthz.fig}. In the case of ESO 358G-29
a bright thick disk is seen, with scale length larger than the thin
disk. This fact that the scale lengths are different causes the
apparent increase of the galaxy's scale height with radius. In the
other early-type spirals, except 1, the thick disk cannot be
seen so clearly, although the scale height here increases with radius. 
We argue that the increase of scale height with radius is linked to the
presence of a thick disk, and with galaxy type, since both effects only
occur in early-type spirals. Based on Fig.~\ref{slopes.fig} it could be
that all spirals of type earlier than 2 have a thick disk with scale
length larger than the thin disk. To check this we have decomposed all
galaxies with type $< 0.0$ into 2 components with constant scale height.
We obtain good fits for all galaxies except for ESO 321G-10, with scale
height ratios ranging from 1.8 to 4.6.

\section{Discussion}

Previous studies have always resulted in confirmations of the assumed
constancy of the vertical scale height, although in some cases authors
refer to the fact that the outermost profiles in their sample galaxies
indicate smaller slopes and hence larger scale heights, e.g. van der
Kruit \& Searle (1981a) for the late-type edge-on galaxies NGC 4244 and 
NGC 5907 and de Grijs \& van der Kruit (1996) for a number of their
earlier-type sample galaxies.

Barnaby \& Thronson, Jr. (1992) find for NGC 5907 that the scale height
is lower between $-100''$ and $+100''$ than in the rest of the galaxy.
They say that this is due to bulge contamination. However, as can be
seen from their Fig. 4, between $\pm 50''$ and $\pm 100''$ the bulge
contribution is negligible.

Van der Kruit \& Searle (1981a) explained these observations by invoking
a possible thickening of the disk just before the radial cut-off they
observed or by attributing it to the influence of an optical warp. 
However, even if the shallowing at the last profiles were entirely due
to a thickening of the disk the implied change in $z_0$ would be
relatively small and restricted to no more than 10\% of the observable
extent of the galaxies.  It should be noted that in a number of profiles
their model deviates significantly from the data, although they
calculate a single scale height and fit all profiles with a constant
vertical scale parameter. 

In general, thick disks seem to occur more often in early-type galaxies
than in later types (e.g., Burstein, 1979).  We have found that the
exponential vertical scale height increases as function of position
along the major axis and that this dependence is strongest for the
earliest-type galaxies.  It seems therefore a natural theory to link
this increase of scale height with radius to the presence of a thick
disk.

For that reason, it is likely that the formation mechanism for the thick
disk and the origin of the increasing scale height effect are similar. 
The possibility that the thick disk is the intermediate component in the
hierarchical formation scenario between the bulge and the thin disk (as
suggested by, e.g., Gilmore, 1984) seems to be ruled out by the large
scale length of the thick disk.  Therefore, more likely formation
mechanisms for the thick disk invoke the accretion of material by the
early thin disk, causing violent dynamical heating processes to take
place, thereby puffing up the thin disk (e.g., Norris, 1987; Statler,
1989). 

Quinn et al. (1993) find that disks following accretions are noticeably
flared. The projected and deprojected scale heights of both the
original disk and of the final system increase with radius. Quinn et
al. (1993) have also shown that the disk is heated in all three
directions by the simulated merger event and that in the outer regions
the radial velocity dispersion is nearly constant, while the vertical
dispersion rises. Van der Kruit \& Freeman (1986) and Bottema (1993)
have studied a number of disk galaxies, especially of later type, in
detail, and found that the velocity dispersions as a function of radius
are consistent with a constant {\it Q} parameter (Toomre, 1964), i.e. 
their measurements are consistent with an exponentially decreasing
vertical velocity dispersion, and assuming a flat rotation curve. 

Bertola et al. (1995) and Fisher (1997) find for their samples of
early-type disk galaxies that the stellar velocity dispersions may
decrease less rapidly than exponentially, along both the major and minor
axes. This result is independent of viewing angle and would indicate
that the radial and vertical velocity dispersions become more nearly
constant with radius. 

These results could imply that our observed increasing scale height with
galactocentric distance may be related to the stellar velocity
dispersion falling more slowly than predicted by the constant {\it Q}
model. Similarly, a non-exponential behaviour of the stellar velocity
dispersion is suggestive of a non-constant disk scale height, or an
increasing thick disk contribution to the mass profile.


\section{Conclusions}

\begin{itemize}

\item We have found that, although constant in a first order
approximation the exponential vertical scale height increases as a
function of position along the major axis and this effect is strongest
for early-type disk galaxies, where this increase can be as much as a
factor of 1.5 per scale length. 

\item This effect might be closely related to the presence of thick
disks, with a scale length larger than that of the thin disk. The fact
that, in this case, the scale length of the thick disk has to be larger
than that of the thin disk strengthens the possibility that the thick
disks were made by merging of small satellites (see, e.g., Quinn et al.,
1993).

\item If, as theory predicts, the scale height is closely coupled to the
vertical velocity dispersion, one should see a less rapid decline of the
vertical velocity dispersion than an exponential decline.

\end{itemize}

\vspace{-0.3cm}
\paragraph{Acknowledgements} - We thank Piet van der Kruit, Roelof
Bottema, David Fisher, Linda Sparke and the referee for the useful
discussions and suggestions concerning this work.

\end{document}